\documentclass[twocolumn,twoside,slac_two,nogroupedaddress]{revtex4}
\usepackage{graphicx}
\usepackage{fancyhdr}
\begin{document}

\title{TeV sources analysis with AGILE}

%

%
\author{Longo, F.$^{1,2}$, Rappoldi, A.$^3$, Lucarelli, F.$^{4,5}$, Pittori, C.$^{4,5}$, Verrecchia, F.$^{4,6}$, Bulgarelli, A.$^7$, Tavani, M.$^{8,9,10}$, on behalf of the AGILE collaboration}
\affiliation{$1$ Dip. di Fisica, Universita' di Trieste, Trieste, Italy}
\affiliation{$2$ INFN, Trieste, Italy}
\affiliation{$3$ INFN, Pavia, Italy}
\affiliation{$4$ ASI Science Data Centre (ASDC), Frascati, Italy}
\affiliation{$5$ INAF-OAR, Astronomical Observatory of Rome, Monte Porzio Catone, Italy}
\affiliation{$6$ Consorzio Interuniversitario di Fisica Spaziale (CIFS), Torino, Italy}
\affiliation{$7$ IASF-INAF, Bologna, Italy}
\affiliation{$8$ IASF-INAF, Roma, Italy}
\affiliation{$9$ Dip. di Fisica, Universita' di Roma ``Tor Vergata'', Roma, Italy}
\affiliation{$10$ INFN, Roma2, Italy}

\begin{abstract}

AGILE (Astrorivelatore Gamma a Immagini LEggero) is an Italian
Space Agency (ASI) mission launched on April 23rd, 2007. The AGILE-GRID instrument onboard of the AGILE satellite is devoted to gamma-ray astrophysics in the 30 MeV - 50 GeV
energy range. This paper presents the results of a systematic study performed
on the first three years of AGILE-GRID data to search for GeV and
sub-GeV counterparts (or derive flux upper limits) of sources detected 
at very high-energy by TeV experiments. Many of the TeV sources are
still unidentified and deep observations at lower energies are needed
to identify their possible counterparts. 

\end{abstract}

\maketitle

\thispagestyle{fancy}

\section{Introduction}

In the last years, the number of TeV gamma-rays sources has increased
up to more than 100, thanks to the observations made by the new
generation of ground-based Cherenkov telescopes HESS, MAGIC and
VERITAS. The identified sources of VHE photons belong mainly to four 
classes: Active Galactic Nuclei (AGNs), Supernova Remnants (SNRs), 
Pulsar Wind Nebulae (PWNs) and
X-ray binaries (XRBs). More than 80 TeV sources are galactic; a large 
fraction of them do not show a clear counterpart and
still remains unidentified (UNID). Multi-wavelength deep observations of 
the region near the TeV positions are needed to identify the possible
counterparts of the UNID TeV sources as well as to constrain the emission
mechanisms of the VHE $\gamma$-ray photons. In what follows, we will
present the results of a search of MeV-GeV counterparts of known and 
UNID TeV sources using the data taken by the AGILE $\gamma$-ray satellite. 
AGILE (Astrorivelatore Gamma ad Immagini LEggero) \cite{tavani1}, \cite{tavani2} is a 
scientific mission of the Italian Space Agency (ASI) dedicated to the observation 
of astrophysical sources of high-energy $\gamma$-rays in the 30~MeV-50~GeV energy range, with
simultaneous X-ray imaging capability in the 18-60~keV band. AGILE is the first 
high-energy mission which makes use of a silicon detector for
the $\gamma$-rays-to-pairs conversion. The AGILE payload combines for
the first time two coaxial instruments: the Gamma-Ray Imaging Detector
GRID (composed by a 12-planes Silicon-Tungsten tracker~\cite{st}, 
a Cesium-Iodide mini-calorimeter \cite{Labanti} 
and the anti-coincidence shield \cite{Perotti}) and the hard X-ray detector Super-AGILE
\cite{SuperAGILE}. The use of the silicon technology allows to have good
performances for the $\gamma$-rays GRID imager in a relatively small and
compact instrument: an effective area of the order of 500~cm$^2$ at
several hundred MeV, an angular resolution of around
3.5$^\circ$ at 100 MeV, decreasing below 1$^\circ$ above 1~GeV, a very
large field of view ($\sim$2.5 sr) as well as accurate timing, positional 
and attitude information. 
AGILE was launched on April 23rd, 2007, and placed in a low ($\sim 550$ km)
equatorial orbit with very small background contamination. 
The AGILE Ground Segment has developed a very efficient alert system, which
leads to automatic alerts sent via email (and sms) 
within $\sim$ 2-2.5 hours after the from an astrophysical event.
Until October 2009, 
the AGILE spacecraft was operated in ``fixed-pointing'' mode,
completing 101 pointings. In Nov. 2009, the spacecraft has to be
re-configured in the safe ``spinning operation mode'', converting
AGILE in an all-sky $\gamma$-ray monitor. The GRID boresight rotation axis now
scans the sky with an angular velocity of about 1$^\circ$/s, accessing about
70-80\% of it each day. AGILE good sensitivity in the 100-400~MeV energy range makes it very
suitable to perform the MeV-GeV counterpart search of TeV sources
described here. 

\section{Selected AGILE results on TeV sources}
\label{section0}

Several TeV emitting sources were studied by AGILE-GRID. Remarkable results were obtained for both Galactic and Extragalactic TeV emitting objects.


\subsection{SNR IC 443} 

The analysis of the intermediate-age Supernova Remnant (SNR) IC 443\cite{ic443} 
revealed a distinct pattern of diffuse emission in the
energy range 100 MeV-3 GeV is detected across the SNR with 
its prominent maximum localized in the Northeastern shell, consistent with the site
of the strongest shock interaction between the SNR blast wave and the dense
circumstellar medium. The location of the High Energy gamma-ray emission is not coincident with the TeV source located
0.4 degrees away, and associated with a dense molecular cloud complex in the SNR
central region. This provides the evidence that electrons cannot
be the main emitters of gamma-rays in the range 0.1-10 GeV at the site of the
strongest SNR shock. The intensity, spectral characteristics, and location of the
most prominent gamma-ray emission together with the absence of co-spatial detectable
TeV emission are consistent only with a hadronic model of cosmic-ray
acceleration in the SNR. This SNR was also studied by Fermi-LAT\cite{FermiIC443}.


\subsection{W28} 

Molecular clouds associated with SNRs can produce gamma-ray emission by means of the interaction of accelerated particles with the concentrated gas. The middle-aged SNR W28, because of its associated system of dense molecular clouds, provides an excellent opportunity to test this hypothesis. AGILE/GRID observations of SNR W28, and their comparisons with observations at other wavelengths (TeV and $^{12}$CO (J=1$\rightarrow$0) molecular line emission) support a hadronic model for the gamma-ray production in W28\cite{giulianiw28}. The AGILE gamma-ray source is positionally well correlated with the TeV emission observed by the HESS telescope. The local variations in the GeV to TeV flux ratio imply that there is a difference between the CR spectra of the north-west and south molecular cloud complexes. A model based on a hadronic-induced interaction and diffusion with two molecular clouds at different distances from the W28 shell can explain both the morphological and spectral features observed by both AGILE in the MeV-GeV energy range and the HESS telescope in the TeV energy range\cite{giulianiw28}. A similar study was performed by Fermi-LAT\cite{FermiW28}. 


\subsection{The Crab Nebula}

The well known Crab Nebula is at the center of the SN1054 supernova remnant. It consists of a rotationally-powered pulsar interacting with a surrounding nebula through a relativistic particle wind. The emissions originating from the pulsar and nebula have been considered to be essentially stable. The surprising detection of strong gamma-ray (100 MeV-10 GeV) flares observed by the AGILE satellite in September, 2010 and October, 2007 were reported by AGILE\cite{Crab}. In both cases, the unpulsed flux increased by a factor of 3 compared to the non-flaring flux. The flare luminosity and short timescale favor an origin near the pulsar, challenging standard models of nebular emission and require power-law acceleration by shock-driven plasma wave turbulence within a ~1-day timescale. The flaring activity of the Crab Nebula was independently confirmed by Fermi-LAT\cite{FermiCrab}. 


\subsection{Vela-X}

Pulsars are known to power winds of relativistic particles that can produce bright nebulae by interacting with the surrounding medium. These pulsar wind nebulae (PWNe) are observed in the radio, optical, x-rays and, in several cases, also in gamma-rays up to TeV energies. The information in the gamma-ray band are useful to reach a comprehensive multiwavelength picture of their phenomenology
and emission mechanisms. AGILE detected the Vela pulsar wind nebula in the energy range from 100 MeV
to 3 GeV\cite{VelaX}. The brightest AGILE gamma-ray source, AGL J0834-4539 has an asymmetric shape incompatible with the AGILE
point-spread function. It is however positionally concident with HESS J0835-455, the TeV source that is identified with the Vela X nebula, implying that AGL J0834-4539 is associated with the pulsar’s PWN.
In the frame of leptonic models, the AGILE measurements are not consistent with a simple
multiwavelength spectral energy distribution involving a single electron population requiring 
additional electron populations to explain the observed GeV fluxes. 
The observations of Vela-X by Fermi-LAT are, however, in favour of a different interpretation\cite{VelaXAbdo}.


\subsection{Mkn 421}

AGILE detected a bright flare from the high-frequency peaked blazar Mrk 421 on June 9–15, 2008\cite{Donnarumma}. 
This flaring state is brighter than the average flux observed by EGRET by a factor of $\sim 3$, but still consistent with the highest EGRET flux.
In hard X-rays (20-60 keV) SuperAGILE resolved a 5-day flare (June 9-15) peaking at $\sim 55$ mCrab.
A large set of simultaneous data, covering a twelve-decade
spectral range was collected for this flare\cite{Donnarumma}, allowing for a deep analysis of the spectral energy distribution as well as of correlated light curves. The $\gamma$-ray flare was interpreted within the framework of the synchrotron self-Compton
model in terms of a rapid acceleration of leptons in the jet.

\section{Systematic search of AGILE counterparts of TeV Sources in the
  AGILE-GRID fixed-pointing data}
\label{section1}

The analysis presented in this section has been performed with an automatic
procedure for all the sources contained in the web-based catalog TeVCat \cite{TeVCat}.
At present, the catalog contains more than 100 TeV sources published
on refereed journals or newly announced by means of Astronomers Telegrams.

The TeVCat is constantly updated with new TeVsources being detected by various TeV experiments. 
TeVCat lists also the basic properties of the different sources such as location and possible association. 
Currently some source classes have been clearly identified, such as extragalactic objects (Blazars and RadioGalaxies), Supernovae Remnants,
Pulsar Wind Nebulae, Gamma-ray Binaries and Dark Sources\cite{TeVCat}.

Our analysis is currently being performed in a systematic way
on a sample of 116 TeV sources, both galactic and extragalactic,
whom centroid positions and extensions have been carefully reviewed
using data from literature. An improved catalog of TeV sources will be
soon available at the ASI Science Data Center (ASDC) webpages \cite{carosi}.  

An automatic procedure, described in \cite{Rappoldi}, search for
possible AGILE counterparts to known TeV sources using the AGILE official
data archive provided by the AGILE Data center (ADC) hosted at the ASDC. 

The analysis has been performed on the full pointing period of
AGILE-GRID (2007.07.09$-$2009.10.31) for a total of 2.3 years of
data. The relevant maps (counts, exposure and diffuse background)
have been generated with an energy threshold of $E_{\gamma} >100$
MeV. All the maps are centered on the selected TeV source
position, with a size of $40^\circ \times  40^\circ$, binned at
$0.1 ^\circ$. The count maps are generated using the event filter named
FM, characterized by the best signal-to-noise ratio.

The source detection step is performed by means of a multi-source
Maximum Likelihood (ML) algorithm (\cite{Mattox}, \cite{bulga}),
fixing the position and the flux of the known AGILE sources from the 
First AGILE Catalog (1AGL sources) \cite{pittori} within a radius of 
$20^\circ$ from the TeV source and letting free the position and the flux of
its possible AGILE counterpart within a radius of $1^\circ$.

The preliminary results, reported in table \ref{tablesource}, 
show that around one fourth of TeV AGNs and
around one third of TeV Galactic Sources (SNR, PWN, XRB or
UNID) have a candidate counterpart in the AGILE data.

Refined analysis will be performed on these last classes due to
possible source confusion, particularly in the case of PWN and UNID TeV sources.

\begin{table*}[!t]
\begin{center}
\caption{Preliminary results of the AGILE-GRID systematic search of the TeVCat sources.}
\label{tablesource}
\vspace{0.15cm}
\small
\begin{tabular}{ccc}
\hline
\hline
Source Class & TeV Catalog & Detection (sqrt(TS $>4$) \\
\hline 
AGN (HBL, LBL) & 45 & 10 \\ \hline
Starburst Galaxies & 2 & \\ \hline
PWN & 24 & 8 \\ \hline
SNR & 14 & 5 \\ \hline
XRB & 3 & 1 \\ \hline
UNID & 25 & 6 \\ \hline
Other Galactic & 3 & 1 \\ 
\hline
\hline
\end{tabular}
\end{center}
\end{table*}

\section{Observations of selected UNID TeV sources in spinning mode}
\label{section2}

Since Nov. 2009, AGILE is observing the $\gamma$-ray sky in spinning mode. The preliminary results of the analysis of
the AGILE-GRID data on a sample of TeV sources, mainly still unidentified were presented at the 32$^{\rm th}$ ICRC. 
These sources are part of a Guest Observer observation program
proposed by the authors (Lucarelli et al.) and approved as targets of the 3rd AGILE
Annoucement of Opportunity (AO3)\cite{LucarelliICRC}.

The AO3 proprietary data cover almost one year of observations in
spinning. The expected sensitivity of AGILE-GRID over 1 year of 
spinning mode observations varies from 15$\times 10^{-8}$ to 
$80\times 10^{-8}~ph~cm^{-2}~s^{-1}$, depending on the exposure of the
observed region.

The analysis of the data was made using the AGILE gamma-ray
analysis software, public release AGILE SW 4.0 \cite{bulga}, downloaded
from the ADC webpages. Only gamma-like events have been considered to generate binned maps of
counts, exposure and models of diffuse emission starting from the
cleaned event list and the auxiliary files provided to the Guest
Observers. To reduce the Earth albedo contamination, all $\gamma$-ray events
whose reconstructed directions form angles smaller than 90$^\circ$
with the satellite-Earth vector were rejected. South Atlantic Anomaly
event cuts have been also applied. 

Table \ref{table_wide} summarizes the preliminary results of the analysis of the GRID
data extracted from a region of interest of $15^\circ$  centered at the proposed TeV
sources~\cite{LucarelliICRC}. Seven of the nine targets proposed in the AO3 program have been
analyzed here. A multi-source ML analysis has been applied to estimate 
the significance of the excess around the target position, expressed 
as the square root of the ML Test Statistic (TS). All the AGILE 1AGL
sources \cite{pittori} have been included in the multi-source analysis. The
$\gamma$-ray sources from the 1st year FERMI-LAT Catalog (1FGL) \cite{abdoCat}
have been included only if their 100-300~MeV flux was above the AGILE
sensitivity over one year.

The table also shows the estimated number of excess counts with respect to the galactic
diffuse emission and the average flux for energies above 100 MeV. The
flux is replaced by the 95\% C.L. upper limits when the $\sqrt{TS}$ of
the ML analysis is below 3. The last column shows the displacement in degree of 
the AGILE excess peak position from the best-fit position of the TeV
emission, as provided from the literature\footnote{The Galactic positions of the
TeV targets have been taken from the Catalog
of TeV Sources in preparation at ASDC \cite{carosi}}. 

\begin{table*}[!t]
\begin{center}
\caption{Preliminary results of the AGILE-GRID observations of
  selected UNID TeV sources.}
\label{table_wide}
\vspace{0.15cm}
\small
\begin{tabular}{cccccccc}
\hline
\hline
TeV source & Date Interval & MJD & Exposure & $\sqrt{TS}$~$^{a}$ & Counts & Flux~$^{b}$ (E$>$100~MeV) & Shift~$^{c}$ \\
& & & [$\times~10^{8}~cm{^2}~s$] & & & [$\times~10^{-8}~ph~cm^{-2}~s^{-1}$] & [deg] \\
\hline 
\hline
HESS~J1614-518 & 2010.01.15$-$2010.10.15 & 55211.5$-$55484.5 & 4.3 & 1.3 & $<$135       & $<$32  & $-$   \\
HESS~J1632-478 & 2010.02.05$-$2010.10.15 & 55232.5$-$55484.5 & 3.8 & 2.8 & $<$187 & $<$49 & $-$ \\
HESS~J1702-420 & 2010.02.05$-$2010.10.15 & 55232.5$-$55484.5 & 3.6 & 0   & $<$28      & $<$7.8    & $-$   \\
HESS~J1731-347 & 2010.02.05$-$2010.10.15 & 55232.5$-$55484.5 & 3.6 & 4.3 & 147 $\pm$ 38 & 41 $\pm$ 11 & 0.34  \\
HESS~J1841-055 & 2010.02.05$-$2010.10.31 & 55232.5$-$55500.5 & 3.6 & 5.8 & 229 $\pm$ 44 & 63 $\pm$ 12 & 0.56  \\
HESS~J1843-033 & 2010.02.05$-$2010.10.31 & 55232.5$-$55500.5 & 3.6 & 0.9 & $<$111     & $<$31     & $-$   \\
HESS~J1857+026 & 2010.02.28$-$2010.11.15 & 55255.5$-$55515.5 & 4.0 & 1.3 & $<$139     & $<$35     & $-$   \\  
\hline
& & & & & & & \\
\multicolumn{8}{l}{\bf{Notes}}\\
\multicolumn{8}{l}{$^{a}$Square root of the Maximum Likelihood Test Statistic (TS) representing the statistical significance of the detection \cite{Mattox}.}\\
\multicolumn{8}{l}{$^{b}$Gamma-ray flux for E$>$100~MeV (or 95\% C.L. 
upper limits) estimated applying the multi-source ML analysis on the proposed target.}\\
\multicolumn{8}{l}{$^{c}$Displacement of the AGILE excess peak position from
  the best-fit position of the TeV target.}\\
\end{tabular}
\end{center}
\end{table*}

\section{Conclusions}
This work presented the results of a systematic search of MeV-GeV
counterparts of TeV sources in the first 2.3 years of AGILE-GRID
``fixed-pointing'' data. The results of the analysis found 
significant $\gamma$-ray excesses within 1$^{\circ}$
from the TeV emission centroid for more than one third of the
TeV targets (116 in total). 
Detailed analyis of the most significant
cases is undergoing. 

A large fraction of the TeV sources is still unidentified. 
The observation of a selected sample of UNID TeV sources has been
proposed and accepted for the 3rd AGILE Announcement of
Opportunity. We also have presented here the analysis of these data, taken 
while AGILE was observing in spinning mode. The preliminary results
show significant detections of E$>$100~MeV photons within 1$^{\circ}$
from the target in two of the seven sources observed: HESS~J1731-347
and HESS~J1841-055. In the first case, the TeV source (recently
associated to a shell-type SNR) is well within the AGILE error
position. 
We note that the UNID sources showing a significant detection, within 0.5deg
from the TeV position, in the spinning data, have been also detected in the systematic search using the pointing data.

Spectral analysis of these two cases, which include the whole AGILE
public data, is undergoing and will be presented in dedicated papers
in progress.

{\it Acknowledgements.} The AGILE Mission is funded by the Italian Space Agency (through contract ASI I/089/06/2) with scientific and programmatic participation by the Italian Institute of Astrophysics (INAF) and the Italian Institute of Nuclear Physics (INFN). We acknowledge financial contribution from the agreement ASI-INAF I/009/10/0.

\clearpage

\end{document}